\documentclass[12pt,preprint]{aastex}

\shorttitle{Flares Sweeping across sunspots} \shortauthors{Li \&
Zhang}

\begin{document}

\title{Statistics of Flares Sweeping across Sunspots}

\author{Leping Li and Jun Zhang }

\affil{Key Laboratory of Solar Activity, National Astronomical
Observatories, Chinese Academy of Sciences, Beijing 100012, China;
lepingli;zjun@ourstar.bao.ac.cn}

\begin{abstract}

Flare ribbons are always dynamic, and sometimes sweep across
sunspots. Examining 588 (513 M-class and 75 X-class) flare events
observed by Transition Region and Coronal Explorer (\emph{TRACE})
satellite and \emph{Hinode}/Solar Optical Telescope (SOT) from 1998
May to 2009 May, we choose the event displaying that one of the
flare ribbons completely sweeps across the umbra of a main sunspot
of the corresponding active region, and finally obtain 20 (7 X-class
and 13 M-class) events as our sample. In each event, we define the
main sunspot completely swept across by the flare ribbon as
A-sunspot, and its nearby opposite polarity sunspots, B-sunspot.
Observations show that the A-sunspot is a following polarity sunspot
in 18 events, and displays flux emergence in 13 cases. All the
B-sunspots are relatively simple, exhibiting either one main sunspot
or one main sunspot and several small neighboring sunspots (pores).
In two days prior to the flare occurrence, the A-sunspot rotates in
all the cases, while the B-sunspot, in 19 events. The total rotating
angle of the A-sunspot and B-sunspot is 193$^\circ$ on average, and
the rotating directions, are the same in 12 events. In all cases,
the A-sunspot and B-sunspot manifest shear motions with an average
shearing angle of 28.5$^\circ$, and in 14 cases, the shearing
direction is opposite to the rotating direction of the A-sunspot. We
suggest that the emergence, the rotation and the shear motions of
the A-sunspot and B-sunspot result in the phenomenon that flare
ribbons sweep across sunspots completely.

\end{abstract}


\section{INTRODUCTION}

Solar flares are one of the most spectacular phenomena in solar
physics. They are sudden brightening in the solar atmosphere, and
consist of a number of components including post-flare loops
\citep{li09}, flare ribbons \citep{iso02,asa04,iso07}, etc. The
flare ribbons are located on either side of the magnetic neutral
line, have magnetic polarities opposite to each other, and move
apart during the process of the flare. Based on the CSHKP magnetic
reconnection model \citep{car64,stu66,hir74,kop76}, the flare
ribbons are caused by the precipitation of nonthermal particles
and/or the effect of thermal conduction which are produced after
magnetic field lines reconnect in the corona. At successive
reconnections, the reconnection points (X-points) move upward, and
therefore, newly reconnected field lines have their footpoints
further out than the footpoints of the field lines that have already
reconnected, which leads us to recognize the `apparent' separation
motion of the flare ribbons. The separation of the flare ribbons has
been used to estimate the electric field in the reconnecting current
sheet \citep[e.g.][]{qiu02,asa04} and also the coronal magnetic
field strength and the reconnection rate \citep[e.
g.][]{iso02,li09b}. As we all know, the separation of the flare
ribbons generally stop at the edge of the sunspots. However, in some
flare events, flare ribbons sweep across the main sunspot of the
flare source region completely.

In this Letter, we statistically study the evolution of the sunspots
and the magnetic fields in the source region of 20 flare events.
This investigation will help us understand the storage and release
of the magnetic energy associated with the flares. The criteria for
the data selection and the methods of the data analysis are
described in Section 2. We present the results and a brief
discussion in Section 3.

\section{DATA AND OBSERVATIONS}

The \emph{Transition Region and Coronal Explorer}
\citep[\emph{TRACE};][]{han99} mission explores the dynamics and
evolution of the solar atmosphere from the photosphere to the corona
with high spatial and temporal resolution. It observes the
photosphere (white-light, WL), the transition region (1216, 1550,
and 1600 \AA), and the 1-2 MK corona (171, 195, and 284 \AA). The
\emph{Hinode}/Solar Optical Telescope \citep{tsu08} focuses on the
vector magnetic field in the photosphere, and dynamics of both the
photosphere and chromosphere associated with the magnetic fields.
The \emph{Broadband Filter Imager} (\emph{BFI}) of \emph{SOT}
obtains data in the Ca II H spectral line (397 nm) and G-band (430
nm) with a 2 minutes cadence and a pixel size of 0.108\arcsec.

In this work, we check all the 75 X-class and 513 M-class flare
events\footnote{http://hea-www.harvard.edu/trace/flare\_catalog/index.html}
from May 1998 to May 2009 observed by \emph{TRACE} and
\emph{Hinode}/SOT, and choose the event displaying that one of the
flare ribbons completely sweeps across the umbra of one main sunspot
of the source region. Finally, we get 7 X-class and 13 M-class
events as our sample (see Table 1). For each event, we mainly use
\emph{TRACE} 1600 \AA~and \emph{SOT} Ca II H images to study the
separation of the flare ribbons. \emph{TRACE} WL and \emph{SOT}
G-band images are employed to study the changes of the sunspots, and
\emph{Solar and Heliospheric Observatory} (\emph{SOHO})/Michelson
Doppler Imager \citep{sch95} observations, are used to survey the
evolutions of the magnetic fields in the source regions.

For the 20 events, we define the main sunspot completely swept
across by the flare ribbon as A-sunspot, and its nearby opposite
polarity main sunspots, B-sunspot. In order to investigate the
evolution of the flares and their source regions quantitatively, we
examine a set of parameters including the rotating angle (RA$_{a}$)
and rotating speed (RS$_{a}$) of the A-sunspot, the rotating angle
(RA$_{b}$) and rotating speed (RS$_{b}$) of the B-sunspot, the
shearing angle (A$_{s}$) and shearing speed (S$_{s}$) of the
A-sunspot around the B-sunspot, the swept area (SA) of the A-sunspot
umbra, the sweeping duration (SD) between the beginning of the flare
and the time when the flare ribbon sweeps across the A-sunspot
completely, and the sweeping speed (SS) of the flare ribbon across
the A-sunspot. To describe the sunspot rotations (shear motions)
conveniently, we denote that the clockwise direction is positive,
and the counter-clockwise one, negative.

In order to illustrate how to measure these parameters, an X3.4
class flare event occurred in active region (AR) 10930 on 2006
December 13 is taken as an example, which is studied by many
authors, e.g. \citet{zha07} surveyed the rotation of the flare
source region and concluded that the interaction between the fast
rotating sunspot and the ephemeral regions triggered the flare,
\citet{iso07} studied the flare ribbons and presented that the
G-band flare ribbons on sunspot umbrae showed a sharp leading edge
followed by a diffuse inside, \citet{jing08} used the separation of
the flare ribbons to study the spatial distribution of the magnetic
reconnection, and \citet{guo08} extrapolated the 3-dimensional
magnetic field configuration with the optimization method. Figure 1
shows two \emph{SOT} Ca II H images at the beginning and the
decaying phases of the flare \citep[see also Figure 1 in
][]{jing08}. The solid, dotted, dashed and dash-dotted lines display
the outer edges of the southern flare ribbon at 2:20, 2:30, 2:40,
and 3:40 UT, respectively. It indicates that the southern flare
ribbon swept across the southern sunspots completely. According to
our definition, the southern sunspots are A-sunspot, and the
northern ones, B-sunspot. The A-sunspot are the following polarity
sunspots, and the B-sunspot, consisting of one main sunspot and
several small neighboring pores, the leading polarity ones. The
parameters SD, SA, and SS in this event are 34 minutes, 120.1
Mm$^{2}$, and 15.1 km s$^{-1}$, respectively.

Figures 2(a)-(b) display two \emph{SOT} G-band images which show the
shear motion of the sunspots in AR 10930. The parameter A$_{s}$
represents the shearing angle (denoted by the arrow in Figure 2(b))
of the A-sunspot around the B-sunspot, and reaches 28$^\circ$ in the
two days before the occurrence of the flare. Using a linear
polynomial fit to the data points, we get the average shearing speed
S$_{s}$ to be 0.5$^\circ$ hr$^{-1}$. Figures 2(c)-(d) show the
rotation of the A-sunspot. The parameter RA$_{a}$ (marked by the
arrow in Figure 2(d)) is calculated using the rotation of penumbral
fibrils around the rotating center (see the circles in Figures.
2(c)-2(d)). In two days prior to the flare occurrence, the RA$_{a}$
is $-$169.5$^\circ$, and the rotating speed RS$_{a}$, $-$2.4$^\circ$
hr$^{-1}$. The parameters RA$_{b}$ and RS$_{b}$ are obtained using
the same methods. In this event, RA$_{b}$ and RS$_{b}$ are
11.6$^\circ$ and 0.5$^\circ$ hr$^{-1}$, respectively. Comparing
Figure 2(c) with Figure 2(d), we notice that the A-sunspot exhibits
obvious flux emergence before the occurrence of the flare.

\section{RESULTS AND DISCUSSION}

In this work, nine parameters (RA$_{a}$ and RS$_{a}$ of the
A-sunspot, RA$_{b}$ and RS$_{b}$ of the B-sunspot, A$_{s}$ and
S$_{s}$ of the shear motion, SA of the A-sunspot, SD and SS of the
flare ribbon) are considered to characterize the phenomenon that the
flare ribbon sweeps across the A-sunspot completely, and listed in
Table 1.

Among all the 75 X-class and 513 M-class flare events observed by
\emph{TRACE} and \emph{Hinode}/SOT, 7 X-class and 13 M-class cases
illustrate that one of the flare ribbons completely sweeps across
the A-sunspot, possessing 9.3$\%$ and 2.5$\%$, respectively. In
these 20 cases, the A-sunspot is the following polarity sunspot in
18 (90$\%$) cases, and displays new emergence in 13 (65$\%$) cases.
The A-sunspot rotates in all cases, and the B-sunspot, in 19
(95$\%$) cases. Among these 19 cases, the rotating directions of the
B-sunspot are consistent with those of the A-sunspot in 12 cases,
occupying 63$\%$. In two days prior to the occurrence of the flare,
the total rotating angles of the A-sunspot and B-sunspot in all
events range from 118 to 315$^\circ$, and the average value is
193$^\circ$. The rotating speeds range from 0.2 to 10.6$^\circ$
hr$^{-1}$, with an average value of 3.2$^\circ$ hr$^{-1}$. Except
one flare event in which the B-sunspot appears as a diffusive
network region, 19 cases show that the B-sunspot is relatively
simple, consisting of either a main sunspot or a main sunspot and
several small neighboring sunspots (pores). All the 19 events
exhibit shear motions of the sunspots, and the shearing directions
of 14 (74$\%$) cases are opposite to the rotating directions of the
A-sunspot. The A$_{s}$s are from 7 to 71$^\circ$, with an average
value of 28.5$^\circ$, and the S$_{s}$s, range from 0.2 to
3.3$^\circ$ hr$^{-1}$, with an average value of 0.9$^\circ$
hr$^{-1}$ in two days before the flare occurrence. The SAs, SDs, and
SSs are 12.5-120.1 Mm$^2$, 8-67 minutes, and 1.6-16.8 km s$^{-1}$,
with an average value of 48.4 Mm$^2$, 35.2 minutes, and 6 km
s$^{-1}$, respectively. All the statistical results are listed in
Table 2.

Only 20 (3.4$\%$) flare events among 588 ones display that flare
ribbons sweep across sunspots completely, indicating that this
phenomenon is few and far between. The A-sunspot is the following
polarity sunspot in 90$\%$ cases, showing that it is easier for the
following polarity sunspots than the leading polarity ones to be
completely swept across by the flare ribbons.

Flares are common phenomena and quickly release magnetic energy
stored in the corona in a short time. The manners of storing the
magnetic energy include rotation and shear motion of sunspots, and
magnetic flux emergence, etc. Rotating sunspots have already been
observed for one century \citep{eve10,mal64,gop65}, and
\citet{ste69} suggested that sunspot rotation may be involved with
the buildup of magnetic energy, which later released by a flare.
\citet{reg06} presented that the slow rotation of the sunspot in
NOAA 8210 enables the storage of magnetic energy and allows for the
release of magnetic energy as flares. Recently, more and more
observations provided the evidence that the rotating sunspot is
involved in the large flare activity \citep[e.g.][]{zha07,yan07}. In
our sample, all the flare events are associated with the sunspot
rotation, consistent with the results mentioned above. The rotating
directions of the A-sunspot and B-sunspot are the same in 63$\%$
cases, which is identical with \citet{yan08} who proposed that the
two sunspots with the same rotating direction have higher flare
productivity than those with opposite rotating directions, and this
scenario is easier to store magnetic energy and increase the
helicity of the flux tube. Furthermore, we find shear motions of the
A-sunspot around the B-sunspot in 19 cases, and the shearing
directions of 74$\%$ cases are opposite to the rotating directions
of the A-sunspot. It seems that the two sunspots with the same
rotating direction and the opposite shearing direction are much
easier to accumulate magnetic energy and produce flares.
\citet{bro03} have shown that some sunspots rotate up to 200$^\circ$
about their umbral center, and the corresponding loops in the corona
fan twist and erupt as flares. \citet{ger03} have simulated the
rotation of a pore around a sunspot, and found that when the pore
rotates 180$^\circ$ around the sunspot, the current increases
rapidly as the center of the pore makes contact with the large
sunspot, which could be an explanation of an observed flare. In this
work, the average total rotating angles of the A-sunspot and
B-sunspot is 193$^\circ$ in two days prior to the occurrence of the
flare, which is corresponding to the results of Brown et al. and
Gerrard et al.. \citet{sch08} used non-linear force-free modeling to
show the evolution of the coronal field associated with a rotating
sunspot, and suggested that the flare energy comes from an emerging
twisted flux rope. In this letter, 13 (65$\%$) flare events display
new flux emergence, that is identical with the simulation of
Schrijver et al.. The parameter SA may represent the magnetic flux
included in the flare. We use the average swept area of 48.4 Mm$^2$
and the magnetic field strength of 1500 G to estimate the magnetic
flux involved in a flare, and get a value of 7.3$\times$10$^{20}$
Mx.

The magnetic field structures and sunspot evolution may respond for
the phenomenon that the flare ribbons sweep across the sunspots. In
our study, the B-sunspot is relatively simple, which means that
almost all the magnetic field lines of the A-sunspot may connect
with the B-sunspot. The rotation and shear motions of the sunspots
twist the magnetic field lines above the source region together,
accumulate the magnetic energy, and inject the emerging twisted flux
into the corona, then sigmoids/$\Omega$ loops erupt as flares and
coronal mass ejections \citep{can99,pev02,reg04}. In the process of
the flare, all the twisted field lines of the A-sunspot are
re-arranged by the successive magnetic reconnection. It is why we
observe that the flare ribbon sweeps across the A-sunspot
completely.

According to the statistical results, we use schematic diagrams to
illustrate the evolution of the magnetic field lines of the
A-sunspot. Figure 3(a) shows the magnetic configuration of the AR
two days before the flare, displaying that all the field lines (see
the green and the red lines) of the A-sunspot connect with the
B-sunspot \citep[see also][]{guo08}. In the following two days, the
rotation (see the thin arrow) and shear motion of the sunspots (see
the hollow arrow) twist the field lines of the A-sunspot together,
inject twisted flux into the corona, form complicated magnetic
topology in the AR (see Figure 3(b)), and accumulate the magnetic
energy. When the magnetic energy exceeds a critical value, e.g. the
rotating angle is beyond 180$^\circ$, the twisted field lines
reconnect, the magnetic energy release, and the flare begin. At the
beginning of the flare, the lower field lines reconnect (see the red
and green lines in Figure 3(c)), and then the lower post-flare loops
appear. The footpoints of these post-flare loops form the flare
ribbons (see the blue lines in Figure 3(c)). Therefore, we find that
the flare ribbons first appear near the magnetic neutral lines of
the two sunspots. As the magnetic reconnection continues and the
reconnection points move upward, the flare ribbons separate (see the
arrows in Figure 3(c)) from each other. As a result, we notice that
the flare ribbons sweep the umbrae of the sunspots. Figure 3(d)
shows the reconnected field lines (see the green and red lines) and
the flare ribbons (see the blue lines) in the decaying phase of the
flare. During the flare, all the field lines of the A-sunspot are
involved in the process of the magnetic reconnection. Therefore, we
find that the flare ribbons sweep across the A-sunspot completely.

However, the statistics is one aspect. More observations, simulation
and theoretical study are needed in order to fully understand the
nature of this phenomenon. An analysis of a typical example of
high-resolution images and magnetograms compared with simulations is
planned.

\acknowledgements

The authors are indebted to the \emph{TRACE}, \emph{Hinode} and
\emph{SOHO}/MDI teams for providing the data. The work is supported
by the National Natural Science Foundations of China (G40890161,
10703007, 10873020, 10603008, 40674081, and 10733020), the CAS
Project KJCX2-YW-T04, and the National Basic Research Program of
China under grant G2006CB806303, and Young Researcher Grant of
National Astronomical Observatory, Chinese Academy of Sciences.

\clearpage

\begin{deluxetable}{ccccccccccc}
\tabletypesize{\scriptsize} \tablecolumns{10} \tablewidth{0pc}
\tablecaption{The 20 flare events and their corresponding 9
parameters}
 \tablehead{ \colhead{Date} & \colhead{Flare} & \colhead{RA$_{a}$ ($^\circ$)} &
\colhead{RS$_{a}$ ($^\circ$/hr)} & \colhead{RA$_{b}$ ($^\circ$)} &
\colhead{RS$_{b}$ ($^\circ$/hr)} & \colhead{A$_{s}$} ($^\circ$) &
\colhead{S$_{s}$ ($^\circ$/hr)} & \colhead{SA (Mm$^{2}$)} &
\colhead{SD (min)} & \colhead{SS (km/s)}} \startdata
1998-Aug-23 & M2.2 & 193.4  & 4     & -121.9& -2    & 27.3    & 0.6   & 18.1  & 28    & 2.1 \\
2000-Feb-08 & M1.3 & -43.1  & -4.1  & -98.8 & -9.5  & 17.9    & 1.9   & 35.2  & 28    & 4.7 \\
2000-Jun-10 & M5.2 & -118.1 & -2.4  & -     & -     & -       & -     & 32.6  & 22.6  & 4.4 \\
2000-Jun-23 & M2.6 & -197.5 & -10.6 & 77    & 3.6   & -30.1   & -1.6  & 12.5  & 8.8   & 9.4 \\
2001-Jan-20 & M7.7 & -52.1  & -3    & -129.1& -7.4  & 11.3    & 0.6   & 51.9  & 45    & 2.2\\
2001-Apr-09 & M7.9 & 70.2   & 1.4   & 56.4  & 1.1   & -34.9   & -0.7 & 104.4 & 36.1  & 6.5 \\
2001-Apr-11 & M2.3 & 80.8   & 1.4   & 70.8  & 1.4   & -35     & -0.7  & 64.4  & 37.7  & 3.6 \\
2001-Apr-12 & X2.0 & 158.1  & 3.5   & 50.1  & 1.1   & -31     & -0.7  & 85.9  & 52    & 3.3 \\
2001-May-12 & M3.0 & 91.9   & 4.7   & 29.5  & 1.3   & 71.3    & 3.3   & 13.8  & 50.3  & 9.2 \\
2002-Jul-29 & M4.7 & -122.4 & -2.7  & 69.3  & 1.3   & 36      & 0.9   & 28.8  & 50.4  & 2.6 \\
2003-Mar-18 & X1.5 & 76.7   & 1.7   & 48.3  & 1     & -47.4   & -1    & 37.1  & 25    & 2.7 \\
2003-May-29 & X1.2 & -173.6 & -8.1  & -128.6& -6.1  & -24.2   & -1.1  & 39.1  & 17.4  & 8.5 \\
2004-Apr-06 & M2.4 & 123.7  & 2.6   & 154.9 & 2.8   & -60.9   & -1.1  & 47.5  & 39.2  & 4.1 \\
2004-Jul-20 & M8.6 & 81.4   & 2.2   & 77.4  & 2.2   & -7.1    & -0.2  & 26.4  & 10.3  & 16.8 \\
2005-Jul-09 & M2.8 & -96.6  & -2    & -85.5 & -0.8  & 22.6    & 0.4   & 30.6  & 27.3  & 5.9 \\
2005-Aug-01 & M1.0 & 84.7   & 1.6   & -78   & -1.5  & 16.4    & 0.4   & 31.4  & 67    & 1.6 \\
2005-Sep-09 & X6.2 & 36.6   & 1.2   & 101   & 4.3   & -10.5   & -0.4  & 72.5  & 58.5  & 1.8 \\
2006-Dec-06 & X6.5 & -213.9 & -9.9  & 11.6  & 0.5   & 14.5    & 0.5   & 34.7  & 18.2  & 7.3 \\
2006-Dec-13 & X3.4 & -169.5 & -2.4  & 12.9  & 0.2   & 28      & 0.5   & 120.1 & 34    & 15.1  \\
2006-Dec-14 & X1.5 & -257.4 & -5.1  & 15.6  & 0.3   & 15.4    & 0.3   & 81.1  & 53.4  & 7.5 \\
\enddata

\end{deluxetable}

\begin{deluxetable}{c|c|c|c|c|c|c|c}

\tabletypesize{\scriptsize} \tablecolumns{10} \tablewidth{0pc}

\tablecaption{The statistical results of the 20 flare events in the
sample}

\tablehead{ \colhead{Character} \vline & \colhead{~} &
\colhead{Rotation} & \colhead{~} \vline & \colhead{Shear} &
\colhead{motion} \vline & \colhead{A Flux} \vline & \colhead{A
polarity}}

\startdata Type & A & B & A=B  & S & S$\neq$A & Emergency & Following \\
\hline Numbers & 20 & 19 & 12 & 19 & 14 & 13 & 18 \\ \hline
Percentage  & 100$\%$  & 95$\%$  & 63$\%$  & 95$\%$  & 74$\%$  & 65$\%$  & 90$\%$ \\
\enddata
\tablecomments{A represents the A-sunspot, B, the B-sunspot, and S,
the shear motion. A=B means that the rotating directions of the
A-sunspot and B-sunspot are the same, and S$\neq$A, the shearing
direction is opposite to the rotating direction of the A-sunspot.}
\end{deluxetable}

\clearpage

\begin{figure}
\epsscale{0.6} \plotone{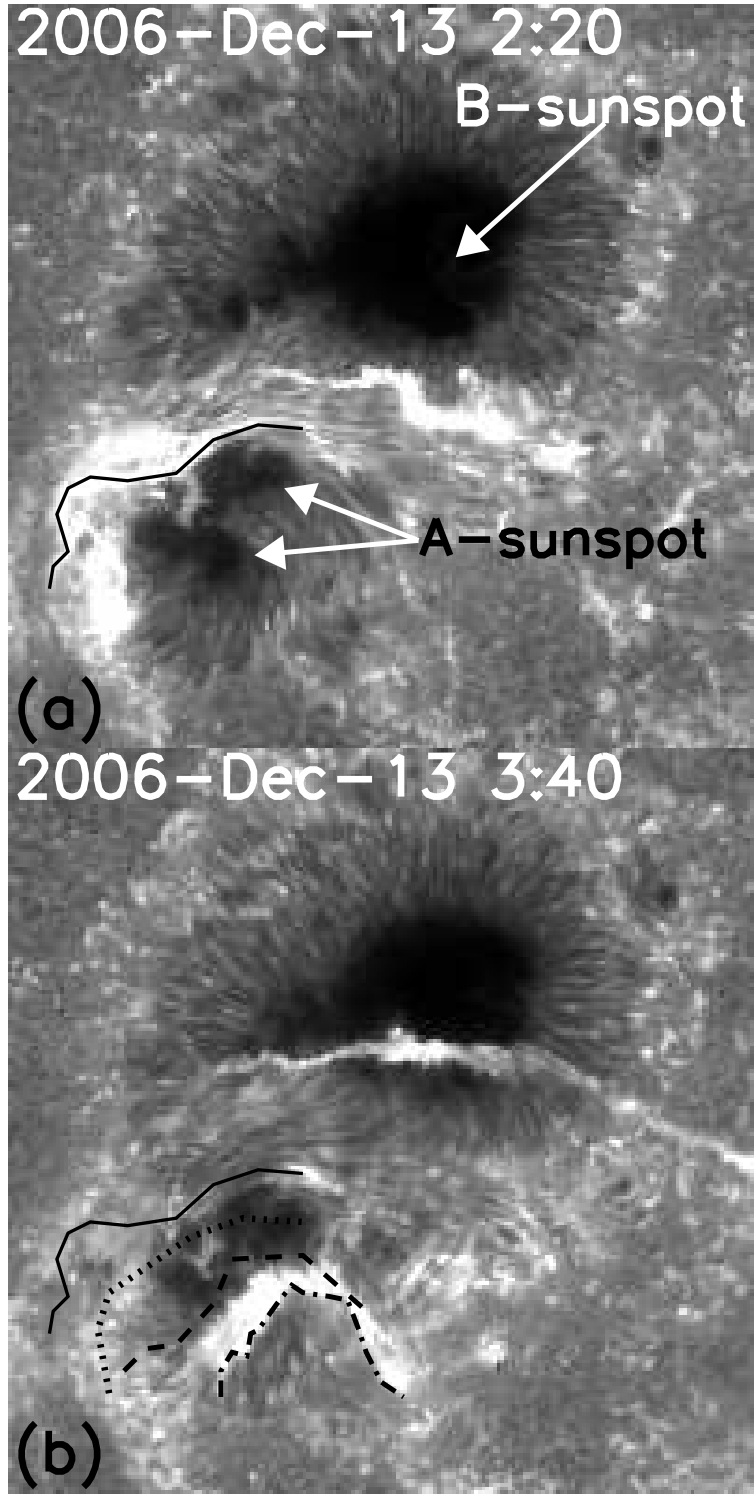} \caption{\emph{Hinode}/SOT Ca II H
images showing the evolution of the flare ribbons. A-sunspot and
B-sunspot mark the sunspots. The curves represent the outer edges of
the southern flare ribbon at different times. The field-of-view
(FOV) is 100\arcsec $\times$ 100\arcsec. \label{f1}}
\end{figure}

\begin{figure}
\epsscale{1.} \plotone{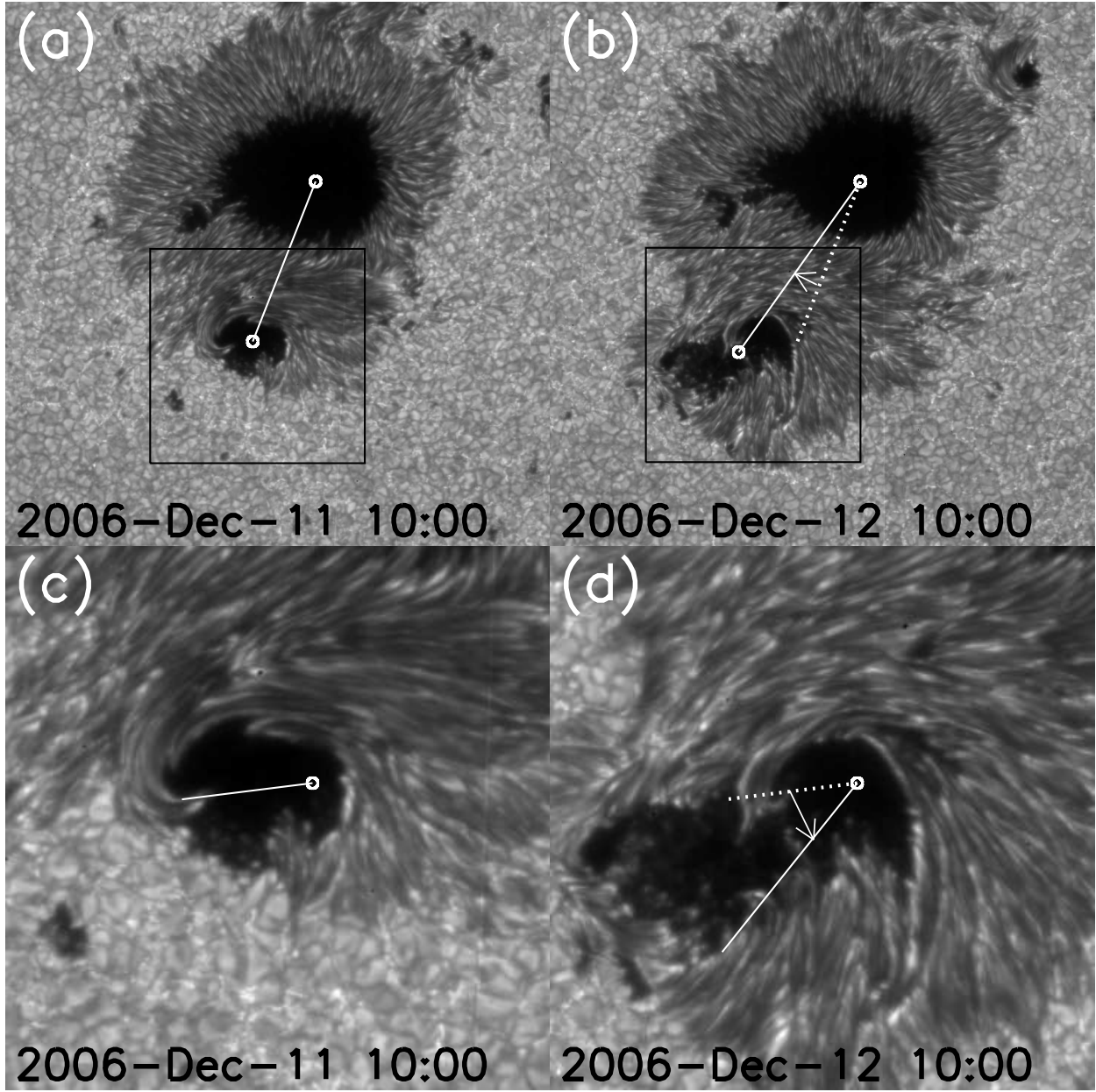} \caption{(a)-(b): Two
\emph{Hinode}/SOT G-band images displaying the shear motion of the
sunspots. The FOV is 100\arcsec $\times$ 100\arcsec. The circles
show the center of the gravity of the A-sunspot and B-sunspot. The
solid lines connect the circles. The dotted line in (b) is a copy of
the solid line in (a). The arrow in (b) shows the shearing angle.
The squares in (a) and (b) denote the FOV in (c) and (d),
respectively. (c)-(d): Two \emph{Hinode}/SOT G-band images showing
the rotation of penumbral fibrils around the rotating center
(circles) \citep[see also][]{zha07}. The solid lines connect the
rotating center and the penumbral fibrils. The dotted line in (d) is
a duplicate of the solid line in (c). The arrow in (d) marks the
rotating angle. The FOV is 40\arcsec $\times$ 40\arcsec. \label{f2}}
\end{figure}

\begin{figure}
\epsscale{0.6} \plotone{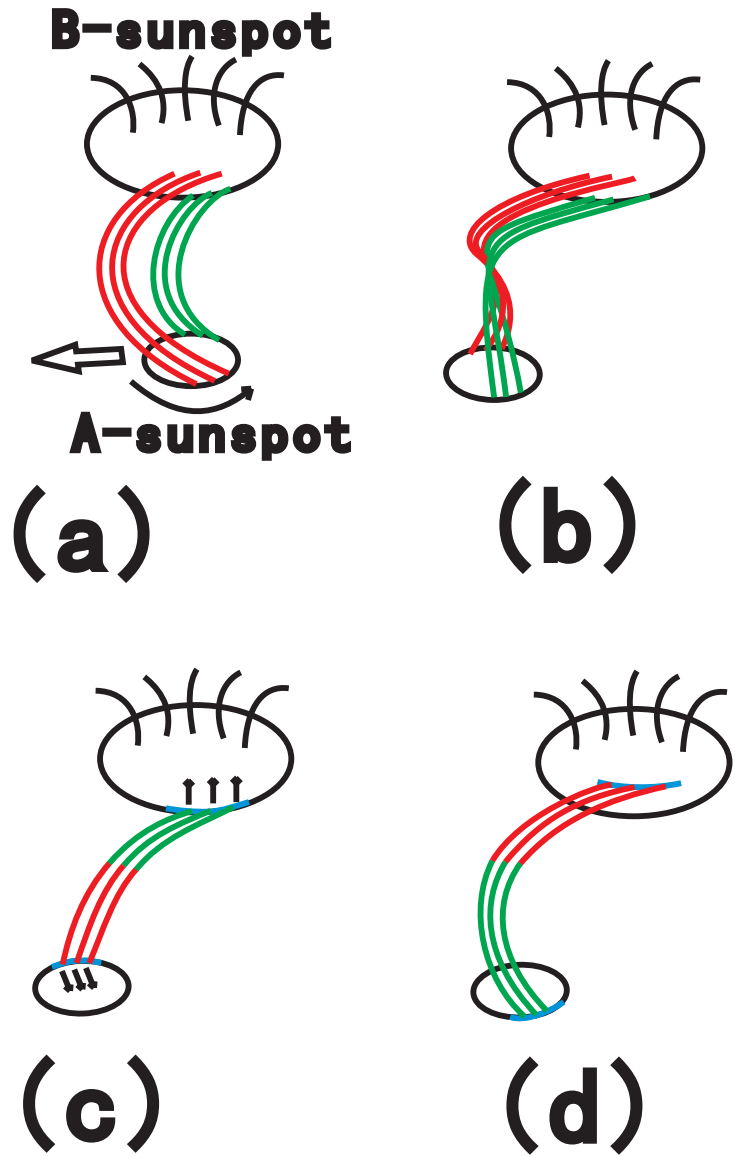} \caption{Schematic diagrams
illustrating the evolution of the magnetic field lines in the source
region of the flare. The ellipses show the sunspots (the A-sunspot
and B-sunspot), and the curves, the magnetic field lines. The thin
arrow in (a) represents the rotating direction, and the hollow one
in (a), the shearing direction. The blue ribbons in (c) and (d)
represent the flare ribbons, and the arrows in (c), the separating
directions of the flare ribbons. \label{f3}}
\end{figure}


\begin{thebibliography}{}
\bibitem[Asai et al. (2004)]{asa04} Asai, A., Yokoyama, T., Shimojo, M., \& Shibata, K.  2004, \apj, 605,
L77
\bibitem[Brown et al. (2003)]{bro03} Brown, D. S., Nightingale,
R. W., Alexander, D., et al. 2003, Sol. Phys., 216, 79
\bibitem[Canfield et al. (1999)]{can99} Canfield, R. C., Hudson, H.
S., \& McKenzie, D. E. 1999, Geophys. Res. Lett., 26, 627
\bibitem[Carmichael (1964)]{car64} Carmichael, H. 1964, in The
Physics of Solar Physics, ed. W. N. Hess (NASA SP-50; Washington,
DC: NASA), 451
\bibitem[Evershed (1910)]{eve10} Evershed, J. 1910, MNRAS, 70, 217
\bibitem[Gerrard et al. (2003)]{ger03} Gerrard, C. L., Brown, D. S.,
Mellor, C., Arber, T. D., \& Hood, A. W. 2003, \solphys, 213, 39
\bibitem[Gopasyuk (1965)]{gop65} Gopasyuk, S. I. 1965, Izv. Krymskoi
Astrofiz. Obs., 33, 100
\bibitem[Guo et al. (2008)]{guo08} Guo, Y., Ding, M. D.,
Wiegelmann, T., \& Li, H. 2008, \apj, 679, 1629
\bibitem[Handy et al. (1999)]{han99} Handy, B. N., et al. 1999,
\solphys, 187, 229
\bibitem[Hirayama (1974)]{hir74} Hirayama, T. 1974, \solphys, 34,
323
\bibitem[Isobe et al. (2002)]{iso02} Isobe, H., et al. 2002, \apj, 566, 528
\bibitem[Isobe et al. (2007)]{iso07} Isobe, H., Kubo, M., Minoshima,
T., et al. 2007, PASJ, 59, S804
\bibitem[Jing et al. (2008)]{jing08} Jing, J.,
Chae, J. C., \& Wang, H. M. 2008, \apj, 672, L73
\bibitem[Kopp \& Pneuman (1976)]{kop76} Kopp, R.
A., \& Pneuman, G. W. 1976, \solphys, 50, 85
\bibitem[Li \& Zhang (2009a)]{li09} Li, L. P., \& Zhang, J. 2009,
\apj, 690, 347
\bibitem[Li \& Zhang (2009b)]{li09b} Li, L. P., \& Zhang, J. 2009,
\apj, 703, 877
\bibitem[Maltby (1964)]{mal64} Maltby, P. 1964, Astrophs. Norvegica,
8, 205
\bibitem[Pevtsov (2002)]{pev02} Pevtsov, A. A. 2002, Sol. Phys.,
207, 111
\bibitem[Qiu et al. (2002)]{qiu02} Qiu, J., Lee, J., Gary, D. E., \& Wang, H. M. 2002, \apj,
565, 1335
\bibitem[R\'{e}gnier \& Amari (2004)]{reg04} R\'{e}gnier, S., \&
Amari, T. 2004, A\&A, 425, 345
\bibitem[R\'{e}gnier \& Canfield (2006)]{reg06} R\'{e}gnier, S., \&
Canfield, R. C. 2006, A\&A, 451, 319
\bibitem[MDI; Scherrer et al. (1995)]{sch95} Scherrer, P. H., et al.
1995, \solphys, 162, 129
\bibitem[Schrijver et al. (2008)]{sch08} Schrijver, C. J., et al. 2008, \apj, 675,
1637
\bibitem[Stenflo (1969)]{ste69} Stenflo, J. O. 1969, \solphys, 8, 115
\bibitem[Sturrock (1966)]{stu66} Sturrock, P. A. 1966, \nat, 211,
695
\bibitem[SOT; Tsuneta et al. (2008)]{tsu08} Tsuneta et al. 2008,
\solphys, 249, 167
\bibitem[Yan \& Qu (2007)]{yan07} Yan, X. L., \& Qu, Z. Q. 2007,
A\&A, 468, 1083
\bibitem[Yan et al. (2008)]{yan08} Yan, X. L., Qu, Z. Q., \& Kong,
D. F. 2008, MNRAS, 391, 1887
\bibitem[Zhang et al. (2007)]{zha07} Zhang, J., Li, L. P., \&
Song, Q. 2007, \apj, 662, L35

\end{thebibliography}
\end{document}